\documentclass[final,3p,times,twocolumn]{elsarticle}

\usepackage{amssymb}
\usepackage{amsmath}
\journal{Chemical Physics Letters}
\usepackage{graphicx}
\usepackage{epstopdf}
\usepackage{epsfig}
\reversemarginpar
\newcommand{\be}{\begin{equation}}
\newcommand{\ee}{\end{equation}}
\newcommand{\benn}{\begin{displaymath}}
\newcommand{\eenn}{\end{displaymath}}
\newcommand{\eb}[1]{\eta_{\rm b} = #1 \times 10^{-5}\, {\rm kg \: m}^{-1}
{\rm s}^{-1}}
\newcommand{\Srbs}{Spontaneous Rayleigh-Brillouin scattering}
\newcommand{\Crbs}{Coherent Rayleigh-Brillouin scattering}

\begin{document}

\begin{frontmatter}



\title{Rayleigh-Brillouin scattering in SF$_6$ in the kinetic regime}


\cortext[mycorrespondingauthor]{Corresponding author}
\author[vua]{Yuanqing Wang}
\author[vua,hua]{Yin Yu}
\author[vua,hua]{Kun Liang\corref{mycorrespondingauthor}}
\ead{liangkun@hust.edu.cn}
\author[cur]{Wilson Marques Jr.}
\author[tue]{Willem van de Water}
\author[vua]{Wim Ubachs}
\address[vua] {Department of Physics and Astronomy, LaserLaB, VU
University, De Boelelaan 1081, 1081 HV Amsterdam, Netherlands}
\address[hua]{School of Electronic Information and Communications,
Huazhong University of Science and Technology, Wuhan, 430074, China}
\address[cur]{Departamento de F\'{\i}sica, Universidade Federal do
Paran\'a, Caixa Postal 10944, 81531-990, Curitiba, Brazil}
\address[tue]{Physics Department, Eindhoven University of Technology,
Postbus 513, 5600 MB Eindhoven, Netherlands}

\begin{abstract}
Rayleigh-Brillouin spectral profiles are measured with a laser-based
scatterometry setup for a 90 degrees scattering angle at a high
signal-to-noise ratio (r.m.s. noise below 0.15 \% w.r.t. peak
intensity) in sulphur-hexafluoride gas for pressures in the range 0.2
-- 5 bar and for a wavelength of $\lambda=403.0$ nm.
The high quality data are compared to a number of light scattering
models in order to address the effects of rotational and vibrational
relaxation.
While the vibrational relaxation rate is so slow that vibration
degrees of freedom remain frozen, rotations relax on time scales
comparable to those of the density fluctuations.  Therefore, the heat
capacity, the thermal conductivity and the bulk viscosity are all
frequency-dependent transport coefficients.
This is relevant for the Tenti model that depends on the values
chosen for these transport coefficients.
This is not the case for the other two models considered: a kinetic
model based on rough-sphere interactions, and a model based on
fluctuating hydrodynamics.
The deviations with the experiment are similar
between the three different models, except for the hydrodynamic model at
pressures $p \lesssim 2\;{\rm bar}$.
As all models are in line with the ideal gas law, we hypothesize the presence of real gas effects in the measured spectra.

\end{abstract}

\begin{keyword}
Rayleigh-Brillouin scattering \sep SF$_6$ gas \sep Tenti model \sep
Rough-sphere scattering model

\PACS 33.20.Fb \sep 51.10+y \sep 51.20+d \sep 47.45.Ab

\end{keyword}

\end{frontmatter}


\section{Introduction}
\label{intro}

Quasi-elastic light scattering is a powerful technique for probing
collisional dynamics and relaxation phenomena in gases and liquids.
Its theory is based in the description of light in terms of
electromagnetism by Lord Rayleigh~\cite{Strutt1899}, and the
scattering process can be explained in terms of fluctuations,
either of density or of entropy. \citet{Brillouin1922} and
\citet{Mandelstam1926} explained that collisions of the gaseous
particles inducing acoustic modes in the medium will affect the line
shape of the scattering spectral profile.
Rayleigh-Brillouin (RB) scattering in various gases has been studied
since the 1960s
~\cite{Greytak1966,Lao1976a,Sandoval1976,Hammond1976,Letamendia1982}
using the method of spontaneous RB scattering, which will be followed
in the present study. Later, alternative methods were developed,
such as stimulated gain Brillouin scattering~~\cite{She1983},
laser-induced gratings~\cite{Stampanoni1995},
coherent RB-scattering~\cite{Grinstead2000,Pan2002} and superheterodyne
optical beating Brillouin spectroscopy~\cite{Tanaka2002,Minami2009}.

At low densities the spectral profile reflects the
Maxwellian velocity distribution of the gas particles.  At larger
pressures density fluctuations also involve collective particle
motion: sound.  The key parameter is the ratio $y$ of the scattered
light wavelength over the mean free path $l_{\rm mfp}$ between
collisions.  More precisely, the uniformity parameter $y$ is defined
in terms of the scattered light wavenumber $k_{\rm sc}$ as $ y = 1 /
k_{\rm sc} l_{\rm mfp}$, and taking for the mean free path the
kinetic approximation $l_{\rm mfp} = v_{\rm th} \eta_s / p$, with the
thermal velocity $v_{\rm th} = (2 k_{\rm B} T / m)^{1/2}$, $\eta_s$ the
shear viscosity, $p$ the pressure, and $m$ the molecular mass.  The
length of the scattering wavevector is
\benn
   k_{\rm sc} = \frac{4\pi n}{\lambda} \: \sin(\theta / 2),
\eenn
with $\lambda$ the wavelength in vacuum, $n$ the refractive index,
and $\theta$ the scattering angle.  At large values of $y$, the density fluctuations are described by the equations of the continuum regime, the Navier-Stokes equations.
At intermediate values of $y$, $y = {\cal O}(1)$, where the
mean free path between collisions is comparable to $1 / k_{\rm sc}$,
light scattering is in the kinetic regime, and density fluctuations
must be described by the Boltzmann equation.  Therefore, whether a
kinetic or a hydrodynamic approach is needed to explain the scattered
light spectrum, not only depends on the density of the gass, but also
on the scattered light wavelength, and thus on the scattering
geometry.

Light scattering not only depends on the translational modes of
motion of a gas, but also on internal degrees of freedom: rotations
and vibrations.  Each mode of motion has its own relaxation rate.
These relaxation frequencies must be compared to a typical frequency
of the density fluctuations: the frequency $f_s$ of sound with
wavelength equal to the scattered wavelength, $f_s = c_s k_{\rm sc} /
2 \pi$, with $c_s$ the speed of sound.  When $f_s$ is much larger
than the relaxation rate of internal degrees of freedom, internal
motion remains frozen on the timescale of the density fluctuations,
and only translational degrees of freedom remain.  Most light
scattering experiments involve translations and rotations only,
unlike experiments at ultrasound frequencies,
where molecular vibrations may come into play.

A kinetic model of scattered light spectra needs information about
the collision properties of the gas molecules, which enters the
collision integral in the Boltzmann equation.  Such information may
be based on explicit molecular interaction parameters.  In one of the
kinetic models discussed in this paper, collisions are between hard
rough spheres that can spin, and thus have internal energy.  These
collisions are parametrized by the hard-sphere radius and the moment
of inertia of the spheres. The roughness allows for the exchange
between translational and rotational energy.

On the other hand for the well-known Tenti
model~\cite{Boley1972,Tenti1974}, the collision integral is
approximated using the known values of the transport coefficients of
a gas. Of these transport coefficients, the heat capacity, the heat
conductivity and the bulk viscosity depend on the relaxation of
internal degrees of freedom.

If transport coefficients are taken as the input of line shape
models, it is important to know if the values used should or should
not allow for the relaxation of internal degrees of motion.  For
example, the bulk viscosity of CO$_2$ used in light scattering with
frequencies $f_s = {\cal O}(10^9)\; {\rm Hz}$ is three orders of
magnitude smaller than the one measured at ultrasound
frequencies~\cite{Pan2005,Gu2014a}.
The reason is that the vibrational relaxation time $\tau_{\rm vib} =
6 \times 10^{-6}\; {\rm s}$ (for a pressure of 1 bar and scaling with
$\propto 1/p$) is comparable to the period of
ultrasound, but slow compared to $1 / f_s$.

Line shape models that take the transport coefficients of macroscopic
gas dynamics as input are attractive because they can be used to
measure these transport coefficients at very large frequencies by
comparing measured scattered light spectra to models.  In this manner
we have recently studied \Srbs-profiles of N$_2$~\cite{Gu2013b} and
CO$_2$~\cite{Gu2014a}, while it was also used to describe
RB-scattering in air, where air was treated as a mono-molecular
species~\cite{Gu2013a,Gu2014b}.  In the same fashion, the bulk
viscosity of several polar and non-polar polyatomic gases was studied
using \Crbs\ \cite{Meijer2010}.
The bulk viscosity is an effective parameter that quantifies the
relaxation of internal degrees of motion in collisions of molecules.
Clearly, the bulk viscosity is frequency dependent, and this
dependency could be measured as a function of frequency in
experiments where light is scattered at different angles.

In this paper we present measurements of Rayleigh scattering
involving sulphur hexafluoride (SF$_6$) molecules.  For SF$_6$ the
vibrational relaxation time is $\tau_{\rm vib} = 2.22 \times
10^{-7}\;{\rm s}$, while the rotational relaxation time is $\tau_{\rm
rot} = 6 \times 10^{-10}\;{\rm s}$ (both values defined for pressures
of 1 bar)~\cite{Haebel1968}.  In our scattering geometry, the typical
Doppler and Brillouin shifts are $f_s = {\cal O}(10^9\; {\rm Hz})$
so that the time scale of measured density fluctuations is approximately two orders of
magnitude larger than $\tau_{\rm vib}$, but it is of the same order
as $\tau_{\rm rot}$.
Therefore, vibrational modes of motion can be ignored, and the
high-frequency values of the transport coefficients should be chosen.
In addition, rotational relaxation may only be partial at the sound
frequencies in our experiment.

In this particular case, models that do not depend explicitly on
values of the transport coefficients may perform better.  In this
paper we will compare two such models to measured SF$_6$ spectra.
\citet{Marques1999} has designed a kinetic theory based on a
rough-sphere interaction model between rotations and translations. As
parameters it takes the moment of inertia of the molecule and the
momentum relaxation rate $\sigma = \eta_s / p$, with $\eta_s$ the shear
viscosity and $p$ the pressure. It replaces the linearized collision
operator with a simple relaxation term $\propto \sigma (f - f_r)$,
where the reference distribution function $f_r$ is designed such that
collisional transfers of momentum and energy agree with those of the
full Boltzmann collision operator. The SF$_6$ molecule is ideally
suited for application of this model as it is nearly spherically
symmetric, while with frozen vibrations its heat capacity ratio $c_p/c_v$ is
close to 4/3, $c_p$ is the heat capacity at constant pressure and $c_v$
is the heat capacity at constant volume.

The other theory by \citet{Hammond1976} is based on fluctuating
hydrodynamics, explicitly involves rotational (and vibrational)
relaxation, and takes measured relaxation rates, $\tau_{\rm
vib}^{-1}$ and $\tau_{\rm rot}^{-1}$ and as input.
%
In order to accommodate more rarified gases, Burnett terms are added
to the continuum equations~\cite{Hammond1976}, but the applicability is still restricted
to values of the uniformity parameter $y \gtrsim 1$. However, for
light scattering off a CH$_4$ gas at $y = 2.70$, the two models, one
kinetic, and the other one hydrodynamic, are hardly distinguishable
\cite{Marques1999}.

The importance of the frequency dependence of the transport
coefficients in explaining light scattering spectra of SF$_6$ was
already recognized by~\citet{Clark1972}, who compares experiments at
$2 < y < 55$ with continuum models.
\citet{Lao1976a} propose to change rotational specific heats from
their zero-frequency value $c_{\rm rot}$ to $c_{\rm rot} / ( 1 +
2\pi\:i\:f_s\:\tau_{\rm rot})$, which results in a frequency
dependence of the thermal conductivity $\lambda_t$ through Eucken's
formula. \citet{Weinberg1973} propose transport coefficients which
not only depend on frequency, but also on the wavenumber of sound.
In general, if the product $2\pi\:f_s\:\tau$ is much larger than 1,
the internal degrees of freedom, while relaxing on a time scale
$\tau$, do not partake in light scattering.  For vibrations of SF$_6$
$2\pi\:f_s\:\tau_{\rm vib} \approx 10^3$, while for rotations
$2\pi\:f_s\:\tau_{\rm rot} \approx 4$, where we used the sound
frequency $f_s = 10^9\;{\rm Hz}$.

Using a recently constructed sensitive light scattering
setup~\cite{Gu2012rsi} with operates at 403~nm and a 90$^\circ$
scattering angle we will revisit Clark's experiments, but with a very
high signal-to-noise ratio.  Much as in \citet{Clark1972} we will
address the frequency dependence of transport coefficients, but now
in the context of both kinetic and hydrodynamic models: (a) a
well-known kinetic model ~\cite{Boley1972,Tenti1974} which takes
transport coefficients as input, (b) a kinetic model which takes
collision parameters, and (c) a continuum model that uses known
relaxation times of the internal degrees of freedom.

\section{Experimental}
\label{experiment}

Rayleigh-Brillouin scattering profiles of SF$_6$ gas were measured
with a sensitive RB-scatterometry setup described
previously~\cite{Gu2012rsi}. A cell equipped with Brewster windows is
placed inside a folded optical cavity to enhance the circulating
power effectively used for inducing RB-scattering to 5 Watt. The
incident laser wavelength was set to $\lambda=403.00$
nm~\cite{Gu2014b} for which a narrowband transmission bandpass filter
(Materion, $T=90$\% at $\lambda=403$ nm and $\Delta\lambda=1.0$ nm)
was available to reject most of the Raman-scattered light.
The RB-spectra were recorded by scanning a plano-concave Fabry-Perot
interferometer (FPI) by tuning its piezo voltage. The four sub-modes
supported in the FPI have a free spectral range of $FSR= 7.498$ GHz,
which was calibrated independently by scanning the laser over some 20
full modes of the FPI with mirror spacings of 5 mm. The instrument
linewidth was measured by imposing a reference laser beam to the FPI
yielding $\delta\nu_{\rm{instr}}= 126.7 \pm 3.0$ MHz.

An important parameter for comparing the measured RB-profiles with
theory is that of the scattering angle $\theta$. A first
determination is obtained by measurements on the geometrical lay-out
of the setup, where narrow pinholes are used to determine beam paths
(see Fig. 1 in~\cite{Gu2012rsi}). Subsequently a test RB-scattering
measurement was performed for 1 bar of argon. Analysis of the
spectrum, on the basis of a Tenti code adapted for argon (including
the macroscopic transport coefficients, while neglecting an effect of
a bulk viscosity) yielded a value of the scattering angle
$\theta=89.6 \pm 0.29$ degrees. While care was taken to not rearrange
the alignment of the setup, this value is used throughout the present
study for measuring an analyzing the spectral profiles for SF$_6$.

RB spectroscopic data were collected by scanning the FPI analyzer in
a stepwise fashion at integration times of 1 s for each position. A
full spectrum covering typically 40 RB-peaks in 10,000 data points
was obtained in some 3 hrs. The scanning axis was linearized and
converted into a frequency scale by computerized interpolation and
matching to the FSR value. Finally the consecutive RB-peaks were
overlaid and added to an averaged RB-spectrum, following procedures
as discussed by \citet{Gu2012rsi}.

The data collection rate of 7,000 counts/sec leads to a
noise-to-background ratio of 0.15\% (w.r.t. peak height) for a
typical spectrum recorded at 1 bar. All experiments were conducted at
room temperature.
Eight measurements (I to VIII) were performed in a sequence of
 pressures ranging from 0.2 to 5 bar for which conditions and gas
 transport coefficients listed in Table~\ref{conditions}.

 \begin{table}
 {\caption{\label{conditions} Values for pressure ($P$) and temperature ($T$) for the eight different measurements of RB-scattering in SF$_6$. In the last column the calculated non-uniformity parameter $y$ is given. Other relevant parameters of the present study are the scattering wavelength $\lambda=403.00$ nm and the scattering angle $\theta=89.6^{\rm{o}}$.}
 }
 \begin{center}
 \begin{tabular}{c c c c c c c}
 \hline
   Data  & $P$ (bar)  & $T$ ($^{\rm{o}}$C)&   $y$-parameter    \\
 \hline
   I     &   0.215       &  25.59     &       0.35  \\
   II    &   0.500       &  23.36     &       0.82 \\
   III   &   0.754       &  24.66     &       1.22  \\
   IV    &   1.002       &  23.36     &       1.64  \\
   V     &   2.002       &  23.36     &       3.26  \\
   VI    &   3.002       &  23.87     &       4.88  \\
   VII   &   4.000       &  23.87     &       6.49  \\
   VIII  &   5.017       &  23.36     &       8.14  \\
 \hline
 \end{tabular}
 \end{center}
 \end{table}

\section{Results}
\label{results}

\begin{figure*}
 \begin{center}
 \includegraphics[scale=0.126]{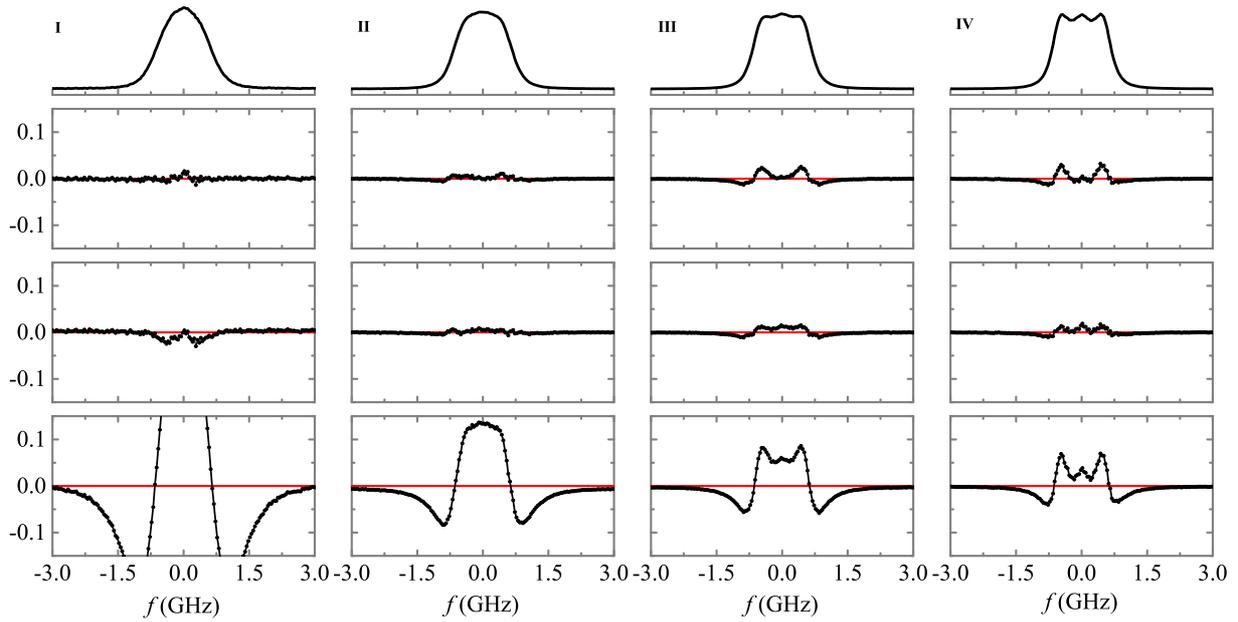}
 \caption{Data on RB-scattering  in SF$_6$, measured under pressure and temperature conditions as listed in Table~\ref{conditions}, indicated by corresponding Roman numerals I-IV, and at $\lambda=403.00$ nm and $\theta=89.6^o$. Top-line: experimental data on a scale of normalized integrated intensity. Second line: deviations of the Tenti-model (S6) description including a frequency dependence of the thermal conductivity and the bulk viscosity $\eta_b$ determined from a least-squares fit. Third line: deviations from a rough-spheres model (with $\kappa=0.227$). In the fourth row, the deviations from the extended hydrodynamic model by Hammond and Wiggins~\cite{Hammond1976} are plotted. Residuals are plotted on a scale of normalized integrated intensity for each profile. }
 \label{fig.resultsA}
 \end{center}
 \end{figure*}

 \begin{figure*}
 \begin{center}
 \includegraphics[scale=0.126]{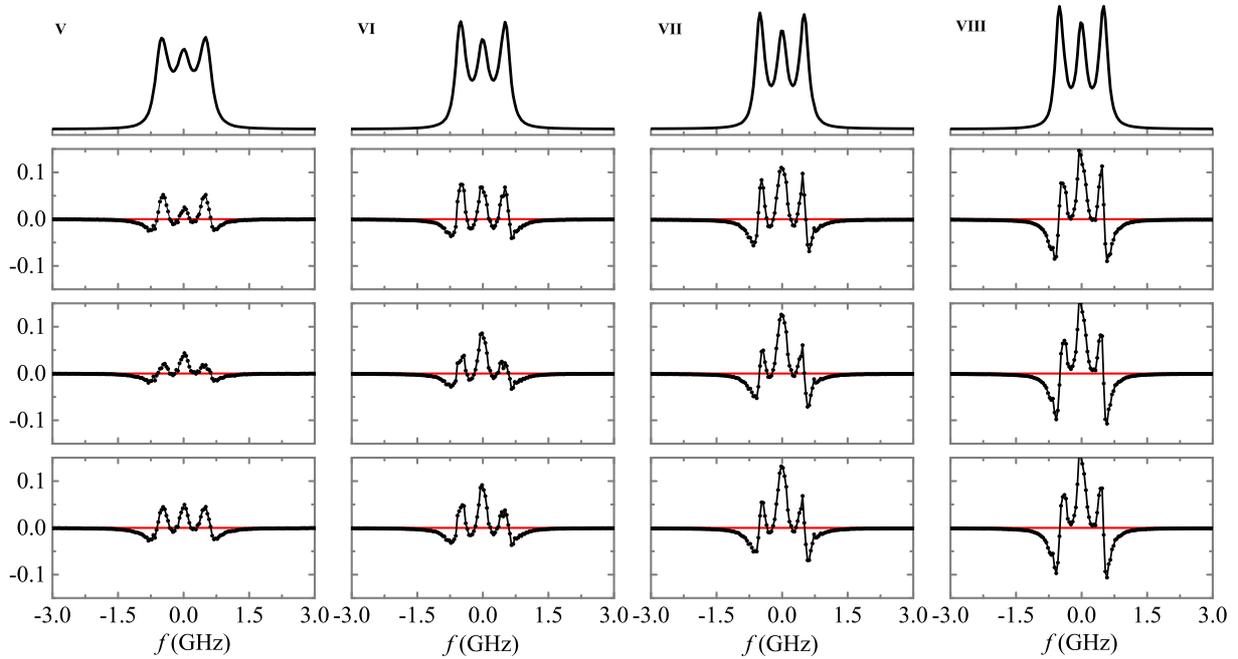}
 \caption{Continued from Fig.~\ref{fig.resultsA} for data sets V-VIII for pressures in the range 2-5 bar. The expression of all the rows are the same as Fig.~\ref{fig.resultsA} (Experimental spectra,  Tenti S6 model with frequency dependent gas coefficients, rough-sphere model, and the extended hydrodynamic model).}
 \label{fig.resultsB}
 \end{center}
 \end{figure*}

The experimental results are shown in Figs.~\ref{fig.resultsA} and
\ref{fig.resultsB} for the eight different experimental pressure conditions
as listed in Table~\ref{conditions}. Due to the larger RB-scattering signals at higher gas pressures,
some form of normalization must be invoked for a consistent comparison between
experiment and model spectra.
For this normalization a scale of equal integrated intensity $I_{\rm{int}}$ over the spectral profile
was chosen. As a result the different peak intensities in the eight experimental spectra, as displayed in the top rows of Figs.~\ref{fig.resultsA} and \ref{fig.resultsB} reflect the unity integration over the spectral profile.
The scaling implies that the large pressure dependence of the RB-scattering of SF$_6$~\cite{Sneep2005} is divided out from the experimental data.
A comparison is made with the results of various models, which will be detailed below.
For this purpose residuals are determined between experimental intensities $I_e(f_i)$
and modelled spectral intensities $I_m(f_i)$, and these are plotted on the same scale of equal integrated
intensity in the second, third and fourth rows in Figs.~\ref{fig.resultsA} and \ref{fig.resultsB}.
We note that we compare in the present study the experimental data with physics-based models,
not relying on ad hoc mathematical functional formulas~\cite{Witschas2011,Ma2014}, to describe the scattering profiles.

As a final figure of merit for the comparison between experiment and the models
a summed and normalized root-mean-square deviation is quantified for the spectral profile:
\begin{equation}
\Sigma_{\rm{nRMS}} = \frac{\sqrt {\frac{1}{N}  \sum_{i=1}^N [I_e(f_i) - I_m(f_i)]^2}} {I_{\rm{int}}}
\label{nRMS}
\end{equation}
which is again normalized to the integrated area of the spectral profile.
Results of such comparisons are displayed in Fig.\ \ref{fig.err} for the eight
different pressure and $y$-parameter conditions and the three models discussed below.

\subsection{Comparison with the Tenti model}
All models need a value of the shear viscosity, for which we took
$\eta_s = 1.52 \times 10^{-5}\, {\rm kg}{\rm m}^{-1} {\rm s}^{-1}$
\cite{Trengove1987,Wilhelm2005}.  For all models we also assumed that
vibrations do not partake in the exchange of translational and
internal degrees of freedom, and that the molecules are spherically
symmetric, so that the heat capacity of internal motion is $c_{\rm
int} = 3/2$.

In addition, the Tenti model needs two possibly frequency-dependent
transport coefficients as input: the thermal conductivity $\lambda_{\rm{th}}$
and the bulk viscosity $\eta_b$.
The zero-frequency value is $\lambda_{\rm th} = 1.30\times 10^{-2}
\;{\rm W} {\rm m}^{-1} {\rm K}^{-1}$ \cite{Kestin1985}, but at the
GHz frequencies of this light scattering experiment the value should
be smaller as vibrational degrees of freedom remain frozen.
At these frequencies we estimate the reduction of $\lambda_{\rm th}$
using Eucken's formula \cite{Chapman1970}
\begin{equation}
 \lambda_{\rm th} = \frac{5}{2} \eta_s c_{\rm t} / m + \rho D (c_{\rm vib} + c_{\rm rot})/m,
\end{equation}
where the heat capacities of translations and vibrations
are $c_{\rm t} = c_{\rm rot} = \frac{3}{2} k_{\rm B}$, and the mass
diffusion coefficient is $\rho D = 20.21 \times 10^{-6}\, {\rm
kg}{\rm m}^{-1} {\rm s}^{-1}$, with $\rho$ the mass density
\cite{Bousheri1987}.  The heat capacity of vibrations $c_{\rm vib} =
7.66 \: k_{\rm B}$ follows from the heat capacity at room temperature
$c_p = 0.664 \; {\rm J} {\rm g}^{-1} {\rm K}^{-1}$ \cite{Guder2009}, which contains
both a rotational and vibrational contribution.  The zero-frequency
value of $\lambda_{\rm th}$ is then reduced by the factor
$(\frac{5}{2} \eta_s c_{\rm t} + \rho D c_{\rm rot}) / (\frac{5}{2}
\eta_s c_{\rm t} + \rho D (c_{\rm vib} + c_{\rm rot}))$,
with the result $\lambda_{\rm th} = 4.72 \times 10^{-3} \;{\rm W}
{\rm m}^{-1} {\rm K}^{-1}$. In view of the approximate character of
Eucken's formula, we ignore a slight pressure dependence of
$\lambda_{\rm th}$ \cite{Kestin1985}.

The kinetic gas model parameters were implemented in an RBS-code for calculating the spectral profiles
within a framework of the  Tenti model and based on the code  by Pan~\cite{Pan2003}.
In the program the bulk viscosity $\eb{1.6}$ was determined in a least squares procedure using the $p = 5\;{\rm bar}$ data. This value for the bulk viscosity was used
in the description of spectral profiles at lower pressures as well. Following a similar
procedure as in \citet{Vieitez2010}, where a value of $\eb{3.5}$ was derived.
We attribute the changed value to an improved instrument resolution,
with respect to the SF$_6$ measurements in previous
experiments~\cite{Vieitez2010}.
The bulk viscosity $\eb{1.6}$ is 300 times smaller than that at low
frequencies \cite{Cramer2012}.

The calculated spectral profiles, obtained with the RBS-code for the Tenti-S6
model implementing frequency-dependent
transport coefficients and  $\eb{1.6}$ are convolved with the instrument width and compared with the experimental
profiles measured for eight different pressures. Residuals of such comparison are displayed in Figs.~\ref{fig.resultsA} and
\ref{fig.resultsB}.

\subsection{Comparison with the rough-sphere model}

Apart from the momentum relaxation rate $\sigma = \eta_s / p$, the
rough-sphere model needs one additional parameter, the dimensionless
moment of inertia $\kappa$ of the SF$_6$ molecule, $\kappa = 4 I / m
d^2$, where $m$, $I$, and $d$ are the mass, moment of inertia and
the effective diameter of the molecule.  The rough-sphere model assumes that the
spherical surfaces of two colliding molecules have no relative
tangential velocity during a collision.  This leads to a solvable
model for the interaction between translational and rotational motion
of the molecules \cite{Chapman1970}.
The moment of inertia $I$ can be assessed directly via spectroscopic
investigation of the rotational level structure of
SF$_6$~\cite{Bobin1987}, or via a direct measurement of the bond
length $R_{SF}$ via X-ray diffraction~\cite{Ewing1963,Bartell1978}, both
yielding a value of $R_{SF}=1.561$ \AA. Via $I=4M_S R_{SF}^2$ this corresponds to a value of
$I = 3.088 \times 10^{-45}$ kg m$^2$ for the moment of inertia for the SF$_6$ molecule.

It should be realized that the effective diameter of the SF$_6$ molecule is not simply $2R_{SF}$
but has a value $d$ which can be determined via experiments on gases.
In experiments measuring the viscosity a value of  $d$ = 4.73 \AA\ was determined~\cite{Mccoubrey1957},
which then results in a value of $\kappa = 0.227$.
Using the rough-sphere relation between shear- and bulk viscosity, $\eta_b = \eta_s \: (6 + 13 \kappa) / 60 \kappa$~\cite{Chapman1970}, a value of the bulk viscosity results: $\eb{1.00}$.
Alternatively, experiments on molecular sieves and zeolites~\cite{Tripp2004,Koenig2012} a slightly larger value for the effective diameter of SF$_6$ molecules was determined, $d$ = 4.90 \AA, yielding a value of $\kappa = 0.211$ and $\eta_b = 1.05 \times 10^{-5}$ kg\,m$^{-1}$s$^{-1}$.
The effect of this difference in effective diameter, resulting in a slightly higher value for
the bulk viscosity, was quantified in a model calculation. It results only in a 0.15\% difference in peak
intensity of the spectral profile, which is negligible.

The formalism used for transforming the rough-sphere model into a RB-scattering spectrum
has been described by Marques~\cite{Marques1999}.
Again, the calculated profile was convolved with the instrument function
and compared with the experimental data in Figs.~\ref{fig.resultsA}
and \ref{fig.resultsB}, with the normalized root-mean-square deviations of the
comparisons presented in Fig.~\ref{fig.err}.

\subsection{Comparison with the hydrodynamic model}
Finally, for the hydrodynamic model by \citet{Hammond1976} we ignored
vibrations, used the rotational relaxation time at $p = 1\;{\rm
bar}$, $\tau_{\rm rot} = 6\times 10^{-10} \; {\rm s}$, and mass
diffusion coefficient
$\rho D = 20.21 \times 10^{-6}\, {\rm kg}{\rm m}^{-1} {\rm
s}^{-1}$~\cite{Bousheri1987}.
The diffusion of rotational energy $D_{\rm rot}$ was assumed equal to
$D$.  The bulk viscosity derived from the rotational relaxation time
is $\eb{1.00}$~\cite{Chapman1970}, while Eucken's relation leads to a
thermal conductivity for translational degrees of freedom
$\lambda_{\rm th \: tr} = 3.25\times 10^{-3} \;{\rm W} {\rm m}^{-1}
{\rm K}^{-1}$.

The hydrodynamic model used is the one described by \citet{Hammond1976}
involving a five-dimensional linear system.  It distinguishes kinetic
and rotational temperatures, with the associated heat flows.
Therefore, it is consistent with a frequency-dependent heat
conductivity \cite{Hammond1976}.
This hydrodynamic model evaluates an equation
\begin{equation}
\partial{\psi(k_{sc},t)}/\partial{t} = -M(k_{sc})\psi(k_{sc},t)
\end{equation}
where $\psi(k_{sc},t)$ spans a 5-component vector with as dimensionless elements the Fourier spatial transforms of the fluctuations of mass density $\bar{\rho}/\rho_0$, translational temperature $\bar{T}/T_0$,
momentum or velocity density $\bar{v}/v_0$, vibrational temperature ${\bar T_{\rm{vib}}}/T_0$, and rotational temperature ${\bar T_{\rm{rot}}}/T_0$.
The $5\times 5$ coefficient matrix $M(k_{sc})$ involves functions of the transport coefficients~\cite{Hammond1976}.
The spectrum of the scattered light is then computed via
\begin{equation}
S(k_{sc},\omega)={\rm Re}[(sI+M(k_{sc}))^{-1}_{11}|_{s=i\omega}]
\end{equation}
evaluating the real part of the (1,1) matrix element of $\psi(k_{sc},s)$, which is the Laplacian of $\psi(k_{sc},t)$.
Hence the scattering profile is derived from the density fluctuations in the medium.
Via this procedure a model spectrum is calculated, which is then convolved with the instrument width and
compared with the experimental spectra in Figs.~\ref{fig.resultsA} and \ref{fig.resultsB}.

\subsection{Difference between experiment and models}

 \begin{figure}
 \centering
 \includegraphics[scale=0.33]{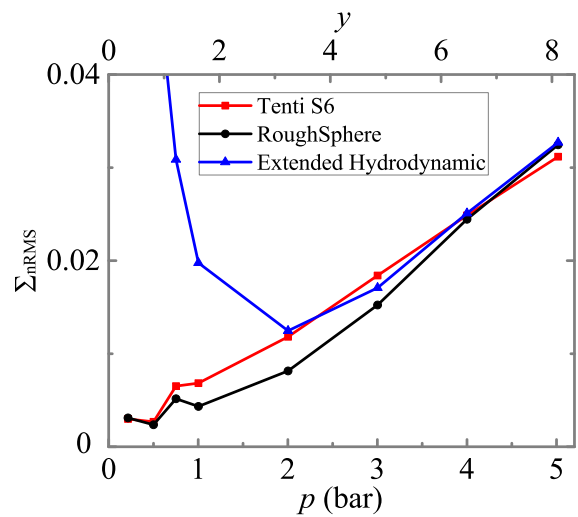}
 \caption{Normalized root-mean square deviations $\Sigma_{\rm{nRMS}}$ as obtained from comparison between experimental spectra and the various line shape models (Tenti-S6, the rough-sphere model,
 and the extended hydrodynamic model), with data points as indicated in the legend. Along the top axis the dimensionless uniformity parameter $y$ is plotted.}
  \label{fig.err}
 \end{figure}

The normalized root-mean-square differences $\Sigma_{\rm{nRMS}}$ between experiment and the three models introduced and discussed in the above is summarized in Fig.~\ref{fig.err}.
At low pressures, corresponding to $y \lesssim 3$, the hydrodynamic model, understandably, fails.
Otherwise the residues of all models with the experiments are very similar: within the experimental accuracy it is impossible to decide in
favor for one of the models.  This is remarkable as the various
models are based on very different physical principles.

However, all models considered share the assumption of SF$_6$ as an
ideal gas.  It is well known that scattered light spectra are
sensitive to real gas effects, i.e. to the deviation from ideal gas law behaviour.
The two kinetic models considered in this paper are derived from the linearized Boltzmann equation, which is consistent with the ideal gas law~\cite{Letamendia1982}.  Similarly, the hydrodynamic model assumes the ideal gas law.
We believe that the residual spectra of Figs.\ \ref{fig.resultsA} and \ref{fig.resultsB}
whose amplitudes increase with increasing pressure, point to real gas
effects.

\section{Conclusion}

Light scattering opens a new window on the properties of SF$_6$ gas, as
relaxation phenomena involve rotational degrees of freedom only, while
acoustic measurements are dominated by vibrational relaxation.
Consequently, the values of the bulk viscosity that we find
$\eb{(1.0-1.6)}$ are 300 times smaller than those at low frequencies
\cite{Cramer2012}.
All three models investigated here to describe the Rayleigh-Brillouin line shape,
indicate that the three rotational modes are involved in light scattering.

A surprising finding is the small but significant residues that for
$y \gtrsim 3$ are very similar for the three models considered.
As the two kinetic models and the hydrodynamic model all assume that
SF$_6$ is an ideal gas, we hypothesize that the deficiency of all
models points to real gas effects.

\section*{Acknowledgement}
Support from the European Space Agency for building the
RB-spectrometer is acknowledged. The authors thank Urs Hollenstein
(ETH Z\"urich) for his help during the calibration of the RBS
scatterometer. YY, KL and YW thank the Chinese Scholarship Council
(CSC) and the Netherlands Universities Foundation for International
Cooperation (NUFFIC) for supporting their stay in Amsterdam and VU
University for the hospitality. KL, WW, and WU acknowledge the
Netherlands Royal Academy of Sciences (KNAW) for funding through the
China Exchange Program.

\bibliographystyle{elsarticle-num-names}

\end{document}